\documentclass[preprint,review,12pt]{elsarticle}
\biboptions{sort&compress}

\usepackage{lineno}

\usepackage{dcolumn}
\usepackage{bm}
\usepackage{multirow}
\usepackage{subfigure}
\usepackage{graphicx}
\usepackage{textgreek}
\usepackage{amssymb}

\newcommand{\degree}{\ensuremath{^\circ}}

\def\mevc  {\ifmmode {\rm MeV}/c \else MeV$/c$\fi}
\def\mevcc {\ifmmode {\rm MeV}/c^2 \else MeV$/c^2$\fi}
\def\gevc  {\ifmmode {\rm GeV}/c \else GeV$/c$\fi}
\def\gevcc {\ifmmode {\rm GeV}/c^2 \else GeV$/c^2$\fi}
\def\ms {\textmu {\rm s}}
\def\O2 {\ifmmode {\rm O}_2 \else O$_2$\fi}
\def\N2 {\ifmmode {\rm N}_2 \else N$_2$\fi}
\def\Rn {\ifmmode ^{222}{\rm Rn} \else $^222$Rn\fi}
\def\Pb210 {\ifmmode ^{210}{\rm Pb} \else $^210$Pb\fi}
\def\Cs {\ifmmode ^{137}{\rm Cs} \else $^137$Cs\fi}

\journal{Astroparticle Physics}

\begin{document}

\begin{frontmatter}

\title {Technical Results from the Surface Run of the LUX Dark Matter Experiment}

\author[cwru]{D.\,S.\,Akerib}

\author[sdsmt]{X.\,Bai}


\author[yale]{E.\,Bernard}

\author[llnl]{A.\,Bernstein}


\author[cwru]{A.\,Bradley}

\author[usd]{D.\,Byram}

\author[yale]{S.\,B.\,Cahn}

\author[cwru]{M.\,C.\,Carmona-Benitez}

\author[brown]{J.\,J.\,Chapman}


\author[cwru]{T.\,Coffey}

\author[umd]{A.\,Dobi}

\author[cwru]{E.\,Dragowsky}

\author[uor]{E.\,Druszkiewicz}

\author[yale]{B.\,Edwards}

\author[brown]{C.\,H.\,Faham}

\author[brown]{S.\,Fiorucci}

\author[brown]{R.\,J.\,Gaitskell}

\author[cwru]{K.\,R.\,Gibson \corref{cor1}}
\cortext[cor1]{Corresponding Author: karen.gibson@case.edu}

\author[ucb]{M.\,Gilchriese}

\author[umd]{C.\,Hall}

\author[sdsmt]{M.\,Hanhardt}

\author[ucb]{M.\,Ihm}

\author[ucb]{R.\,G.\,Jacobsen}

\author[yale]{L.\,Kastens}

\author[llnl]{K.\,Kazkaz}

\author[umd]{R.\,Knoche}

\author[yale]{N.\,Larsen}

\author[cwru]{C.\,Lee}


\author[lbl]{K.\,T.\,Lesko}

\author[lip]{A.\,Lindote}

\author[lip]{M.\,I.\,Lopes}

\author[yale]{A.\,Lyashenko}

\author[brown]{D.\,C.\,Malling}

\author[tamu]{R.\,Mannino}

\author[yale]{D.\ N.\,McKinsey}

\author[usd]{D.\,Mei}

\author[davis]{J.\,Mock}

\author[uor]{M.\,Moongweluwan}

\author[harvard]{M.\,Morii}

\author[ucsb]{H.\,Nelson}

\author[lip]{F.\,Neves}

\author[yale]{J.\,A.\,Nikkel}

\author[brown]{M.\,Pangilinan}

\author[cwru]{K.\,Pech}

\author[cwru]{P.\,Phelps}

\author[tamu]{A.\,Rodionov}

\author[cwru]{T.\,Shutt}

\author[lip]{C.\,Silva}

\author[uor]{W.\,Skulski}

\author[lip]{V.\,N.\,Solovov}

\author[llnl]{P.\,Sorensen}


\author[tamu]{T.\,Stiegler}

\author[davis]{M.\,Sweany}

\author[davis]{M.\,Szydagis}

\author[ucb]{D.\,Taylor}

\author[davis]{M.\,Tripathi}

\author[davis]{S.\,Uvarov}

\author[brown]{J.\,R.\,Verbus}

\author[lip]{L.\,de\,Viveiros}

\author[davis]{N.\,Walsh}

\author[tamu]{R.\,Webb}

\author[tamu]{J.\,T.\,White}

\author[harvard]{M.\,Wlasenko}

\author[uor]{F.\,L.\,H.\,Wolfs}

\author[davis]{M.\,Woods}

\author[usd]{C.\,Zhang}

\address[brown]{Brown University, Dept. of Physics, 182 Hope St., Providence, RI
02912}

\address[cwru]{Case Western Reserve University, Dept. of Physics, 10900 Euclid
Ave, Cleveland, OH 44106}

\address[harvard]{Harvard University, Dept. of Physics, 17 Oxford St., Cambridge,
MA 02138}

\address[lbl]{Lawrence Berkeley National Laboratory, 1 Cyclotron Rd., Berkeley,
CA 94720}

\address[llnl]{Lawrence Livermore National Laboratory, 7000 East Ave., Livermore,
CA 94551}

\address[lip]{LIP-Coimbra, Department of Physics, University of Coimbra, Rua Larga, 3004-516 Coimbra, Portugal}

\address[moscow]{Moscow Engineering Physics Institute, 31 Kashirskoe shosse, Moscow
115409}

\address[sdsmt]{South Dakota School of Mines and Technology, 501 East St Joseph
St., Rapid City, SD 57701}

\address[tamu]{Texas A \& M University, Dept. of Physics, College Station, TX 77843}

\address[ucb]{University of California Berkeley, Dept. of Physics, Berkeley, CA
94720-7300}

\address[davis]{University of California Davis, Dept. of Physics, One Shields Ave.,
Davis, CA 95616}

\address[ucsb]{University of California Santa Barbara, Dept. of Physics, Santa
Barbara, CA 95616}

\address[umd]{University of Maryland, Dept. of Physics, College Park, MD 20742}

\address[uor]{University of Rochester, Dept. of Physics and Astronomy, Rochester,
NY 14627}

\address[usd]{University of South Dakota, Dept. of Physics, 414E Clark St., Vermillion,
SD 57069}

\address[yale]{Yale University, Dept. of Physics, 217 Prospect St., New Haven,
CT 06511}

\begin{abstract}
We present the results of the three-month above-ground commissioning
run of the Large Underground Xenon (LUX) experiment at the Sanford
Underground Research Facility located in Lead, South Dakota, USA.  LUX
is a 370~kg liquid xenon detector that will search for cold dark
matter in the form of Weakly Interacting Massive Particles (WIMPs).
The commissioning run, conducted with the detector immersed in a water
tank, validated the integration of the various sub-systems in
preparation for the underground deployment.  Using the data collected,
we report excellent light collection properties, achieving
8.4~photoelectrons per keV for 662~keV electron recoils without an
applied electric field, measured in the center of the WIMP target.  We
also find good energy and position resolution in relatively
high-energy interactions from a variety of internal and external
sources.  Finally, we have used the commissioning data to tune the
optical properties of our simulation and report updated sensitivity
projections for spin-independent WIMP-nucleon scattering.
\end{abstract}



\end{frontmatter}


\section{Introduction} \label{sec:intro}

The goal of the LUX experiment is to detect or exclude WIMP-nucleon
elastic scattering interactions~\cite{Ref:WIMP} with scalar cross
sections of $7\times 10^{-46}~\mbox{cm}^2$~\cite{Ref:LUX_design} at a
WIMP mass of 100~\gevcc, equivalent to 0.5~events/100~kg/month in the
inner 100~kg fiducial volume of the 370~kg liquid xenon (LXe)
detector.  Events in the LXe target create direct scintillation light
(S1), while electrons escaping recombination at the event site are
drifted to the liquid surface and extracted into the gas phase by
applied electric fields, where they create electroluminescent light
(S2) (see Ref.~\cite{Ref:Araujo} for a comprehensive review of these
processes).  Both S1 and S2 processes emit vacuum-ultraviolet (VUV)
light peaking at 178~nm.  The dominant backgrounds to the WIMP search
consist of nuclear recoils, due to neutrons, and electron recoils,
primarily from external \textgamma-rays and internal \textbeta~decays.
Local external backgrounds are minimized by the appropriate choice of
detector materials, including radio-pure titanium for the cryostat and
low-radioactivity photomultipliers (PMTs) for light readout, and the
inclusion of water shielding around the detector. In addition, two
data analysis methods help to discriminate against the remaining
background: (i) separation of electron recoil interactions from
nuclear recoils, based on the amount of S2~light relative to S1~light;
and (ii) the strong self-shielding capability of the dense LXe that,
when coupled with the use of three-dimensional event reconstruction,
significantly reduces the electromagnetic and neutron backgrounds that
occur primarily in the outermost LXe.  As in most direct search
detectors, WIMP interactions are indistinguishable from single elastic
neutron scatters in LUX, so great care must be taken to minimize the
number of neutrons propagating to the fiducial region of the LXe
chamber.

The LUX~experiment will begin to search for WIMP dark matter in the
Davis campus at the 4850-foot level (1480~m) of the Sanford
Underground Research Facility (SURF) in late 2012. The underground
deployment of such an experiment is a complex process and a major goal
of the surface run was to validate the various sub-systems and to
verify the integration of the entire system.  Some LUX components
involve novel technical solutions that benefited from realistic
testing, including the cooling, gas circulation and purification
systems, the control and safety systems, high voltage delivery, and
the data acquisition system. Conducting the detector commissioning
above ground also allowed corrective actions to be implemented more
effectively prior to underground running. A second aim of the run was
to allow a preliminary assessment of the radiation-detection
performance of the experiment. Besides early validation of key design
parameters, such as the light collection of the chamber, this also
exercised the general data analysis procedures, including the
reconstruction of LXe interactions from a variety of radioactive
sources.  While rigorous characterization of the low-energy
performance of the experiment can only be carried out in the
low-background environment of an underground laboratory, these complex
data analysis procedures require substantial development, and the
early data from this run has proved to be very valuable in advancing
the analysis effort.

This article is organized as follows. We discuss the experimental
set-up for the surface run, the performance of the integrated system,
and surface data-taking in Section~\ref{sec:det}.  Measurements of the
light and charge responses for high-energy background and calibration
sources are presented in Section~\ref{sec:results}.  The outlook for
the LUX dark matter search, based on the results presented here, is
discussed in Section~\ref{sec:summary}.

\section{LUX detector commissioning} \label{sec:det}

\begin{figure}[htbp]
  \centerline{
  \includegraphics[width=1.0\hsize]{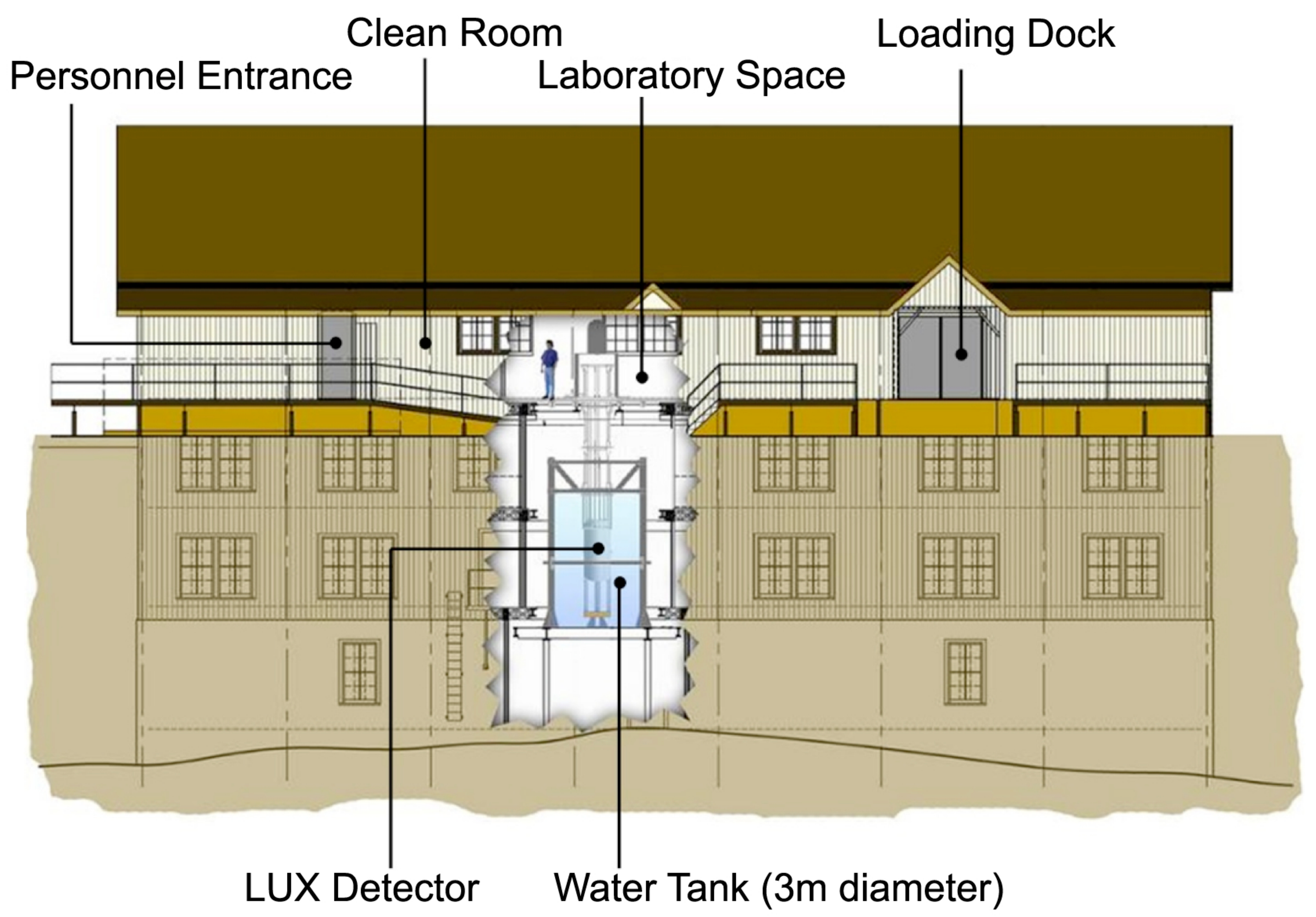} }
  \caption{A cutaway of the surface facility provided by SURF for LUX
  detector commissioning is shown.  All detector systems were tested
  above ground, including detector deployment in a 3~m diameter water
  shield.}  \label{fig:surface_bldg}
\end{figure}

The commissioning of the LUX detector was conducted at a multi-level
surface facility outfitted by SURF for this purpose, depicted
schematically in Fig.~\ref{fig:surface_bldg}.  The detector was
deployed for full commissioning during August 2011; the
system-integrated commissioning began on September 1, 2011, and
extended through February 14, 2012, with more than 100 days of
cryogenic detector operation.  During the surface run, the detector
was filled with 370~kg of xenon, with 300~kg in the active region
between the cathode and anode wire planes.  The surface facility lies
1.6~km above sea level and has higher cosmic-related backgrounds than 
are found at sea level.  In order to test all aspects of the
underground detector deployment and to reduce \textgamma~and neutron
backgrounds during data-taking, the detector was operated within a 3~m
diameter water tank designed for commissioning.  The water tank
provided approximately 1~m of shielding around the cryostat and
reduced the total \textgamma-ray~background, which is largely due to
the concrete and wood building materials used in the surface facility,
from 10~kHz to 100~Hz.  The water shield also reduced the
cosmic-ray-induced neutron background from 240~Hz to 10~Hz.  The
$108.8\pm 0.3$~Hz muon background measured at the surface lab is not
significantly affected by the shield.

The LUX experiment is described in detail in Ref.~\cite{Ref:LUX_NIM}.
The LUX detector is a cylindrical two-phase xenon time-projection
chamber (TPC), with instrumented xenon covering 55~cm in the vertical
($z$) direction and 24~cm in radial extent.  The xenon scintillation
light is detected with 122 VUV Hamamatsu R8778 photomultiplier tubes
(PMTs) evenly divided between top and bottom arrays, operating within
the gas and liquid phases, respectively, and held in place with copper
support structures.  The VUV scintillation light is reflected by
twelve polytetrafluoroethylene (PTFE) panels~\cite{Ref:PTFE} that
fully cover the length of the active region and additional PTFE
reflectors that cover all copper surfaces in the PMT support
structures.  The PTFE panels are 1~cm thick in the radial direction
and are mounted on a polyethylene support structure.  The cryostat is
fully immersed in the water shield and all electrical and gas
connections are made through conduits that emerge from the top of the
cryostat and lead to a cart located on the experimental deck.

To extract ionization from interaction sites and produce S2~light, an
electric field is applied along the $z$-axis of the detector using
four wire planes and an anode wire-mesh plane.  The field between the
cathode and gate wire planes is used to drift the electrons vertically
through the liquid xenon, away from the event site and toward the gas
gap above the gate wire plane (hereafter called the ``drift field'').
The field between the gate wire plane and the anode mesh plane then
extracts the ionization electrons into a layer of gas, approximately
5~mm thick, where they generate S2 light before being collected on the
anode mesh; the terms ``extraction field'' and ``electroluminescence
field'' apply just below and above the liquid surface, respectively.
Additional wire planes are included above the bottom PMT bank and
below the top PMT bank to terminate the field lines and shield the PMT
optics.  Data taken without an applied electric field provides only
S1~light.  ``Dual-phase'' data, which is taken with applied electric
fields, produces S1 light in the liquid xenon and S2 light in the
layer of gaseous xenon between the liquid surface and anode mesh
plane.

\subsection{Detector operation} \label{sec:det_ops}

Detector cooling from room temperature to 185~K was achieved in eleven
days through the use of a liquid nitrogen thermosyphon
system~\cite{Ref:LUX_TS}.  During initial cooling, a maximum cooling
rate of 0.8~K/hr was allowed.  Although the thermosyphons could
deliver a significantly higher rate of cooling, this would risk
establishing thermal gradients that could lead to warping of the PTFE
panels.  Following the initial detector cooling, a stable operating
temperature of approximately 175~K was achieved, although variations
in the detector temperature between mid-November and mid-February
occurred due to operational changes in the xenon circulation.  The
detector pressure was stable within 0.3\% during at least one four-day
period of minimal configuration changes over the New Year holiday.

During surface commissioning, we observed a limitation in the drift
field from the onset of electroluminescent discharge on the cathode
grid wires at 10 kV.  This cathode grid had 10~mm wire spacing and a wire
diameter of 100~$\mu$m.  We had also established a limitation of
the cathode high-voltage feedthrough to 20 kV and a very conservative operating field
of 62~V/cm, corresponding to an electron drift velocity of 1.2~mm/\ms,
was used during dual-phase data taking.  A maximum operating field of
120~V/cm was achieved, but we chose to operate at half of that field
to remain well within a safe range. In order to address these issues, both the cathode wire plane and the
feedthrough were redesigned and new bench-tested versions manufactured
ahead of underground deployment, which should enable LUX to operate at
a nominal field of 800~V/cm.

The purification system is designed to circulate the xenon gas through
a heated zirconium getter, with the help of an external pump, which
requires returning the xenon to the gas phase and re-condensing
it~\cite{Ref:Dobi}. Following the initial liquefaction, it was
discovered that an internal plumbing fitting in the circulation line,
designed to transfer condensed xenon to the bottom of the detector,
came loose during assembly, and the intended xenon circulation path
was compromised during the remainder of the surface commissioning.
Consequently, xenon purification was explored in several circulation
modes over a period of 46~days, including convection and circulating
through a plumbing line intended for liquid recovery.  Using the
original circulation path with the loose fitting, we were able to
circulate xenon through the gas handling system at a rate of 35~slpm,
corresponding to 300~kg/day, with a net heat load $< 5$~W.  During
xenon circulation, the purity of the gas prior to getter purification
was monitored with a cold-trap-enhanced mass spectrometer
technique~\cite{Ref:cold_trap}, and impurity concentrations of 0.4~ppb
O$_2$ and 0.5~ppb N$_2$ were obtained.  The maximum electron drift
length achieved within the chamber, which gives a measure of the
reduction of electronegative impurities, was 25~cm (200~\ms~mean
electron lifetime).  This corresponds to half the length of the active
region between the cathode and gate wire planes.  This measurement is
discussed further in Section~\ref{sec:purity}.

\subsection{Surface data-taking} \label{sec:data_taking}

We collected zero-field data regularly between mid-November 2011, when
we began to condense xenon in the detector, and mid-February 2012,
when the surface commissioning ended to prepare the detector for
transport underground.  Dual-phase data were collected between
mid-December and mid-February.  The data were recorded through the
full data-acquisition (DAQ) chain, providing the opportunity to assess
the performance of the system from beginning to end and to debug all
features of the electronics chain and data-taking that could otherwise
cause pathologies in the dark matter search data.

A detailed description of the data acquisition system can be found in
Ref.~\cite{Ref:LUX_DAQ}.  The analog signal is digitized at an
operational sampling rate of 100~MHz by Struck ADC modules.  Only
candidate pulses that pass above a hardware threshold are digitized
(called ``pulse-only digitization'' or POD) in order to reduce the
recorded event size by a factor of fifteen, while preserving full
sensitivity to dark matter signals.  The POD recorded 24~samples
before the signal threshold for pulse detection was crossed.  We also
recorded an additional 31~trailing samples after the pulse dropped
below a second threshold, which was consistent with the measured
electronics baseline noise and that defined the end of the pulse.  For
most surface data collected, a pulse detection threshold of 1.5~mV was
used.  The efficiency of noise rejection using zero suppression was
measured by setting a threshold for the POD corresponding to a 95\%
efficiency to detect single photoelectrons (phe) at nominal PMT gains,
while biasing the PMTs to only $-100$~V and grounding the wire planes.
This allowed us to test for noise in the electronics chain,
independent of any noise nominally caused by the PMTs being at full
bias voltage.  We obtained a zero suppression efficiency $> 99.999\%$
in this configuration.  During surface operation, the DAQ was able to
handle a 1.7~kHz acquisition rate during dual-phase operation without
any dead time in downloading data.  A maximum event rate of 1.5 kHz 
can be sustained with no dead time even if every optical signal were to 
be recorded by all PMTs, which is a conservative assumption.

The LUX trigger system, described in Ref.~\cite{Ref:LUX_NIM}, was
configured to trigger on either S1 or S2~signals for the surface
data-taking.  The trigger consists of two 8-channel digital signal
processors (DDC-8DSP).  Sixteen trigger groups are defined, with each
group consisting of eight PMTs located in either the top or bottom PMT
array.  Digital filters are used to search for S1 and S2~signals in
the analog sum of the signals in the PMT~groups.  The multiplicity of
the S1 and S2~signals, with pulse areas between a lower and an upper
limit, was used to generate triggers.  All of the data shown in this
paper were collected with an S1 trigger.  Events were defined within a
window of 500~\ms~following the S1 trigger, which allowed the
collection of pulses along the full length of the detector.
Double-triggering was prevented by a trigger hold-off for the
500~\ms~following a trigger.  This mode of operation is different from
that planned for underground operation, which will allow multiple
triggers in an event.

Muons were a significant source of background during the surface
data-taking, with a flux of $0.019\pm
0.003~\mbox{cm}^{-2}\mbox{s}^{-1}$ measured at the surface facility,
which is approximately 14\% higher than the value at sea level.  At the 4850-foot
underground level of SURF, the muon flux is reduced to $(4.4 \pm
0.1)\,\times\,10^{-9}~\mbox{cm}^{-2}\mbox{s}^{-1}$~\cite{Ref:Mei}
($\sim$\,4~muons per day across the active region of the LUX
detector), which corresponds to $4.3 \pm 0.2$~km of water equivalent
shielding.  For our dark matter search, we require single phe
sensitivity in the PMTs and DAQ electronics and both have been
designed to operate in a low background environment, with small
average VUV photon rates generated within the chamber.  Consequently,
the electronics chain was optimized to provide single phe sensitivity
when operating at a PMT gain of $4\,\times\,10^6$, but not at PMT
gains significantly lower than this.  Since cosmic muons can deposit
large amounts of energy in the detector, particularly during
data-taking with both S1 and S2~light, we limited both the PMT gain
and the extraction and electroluminescence fields between the gate and
anode wire planes to safeguard the PMTs during data-taking above
ground. Consequently, the PMT gains were limited to $1\,\times\,10^5$
during dual-phase data taking. The DAQ readout configuration for these
data required a PMT pair above the nominal POD threshold at 1.5~mV.
At the lower gain, this threshold impairs our ability to reconstruct
pulses with areas lower than approximately 20~phe for S1~pulses and
500~phe for S2~pulses and produces a non-trivial threshold effect that
is not readily de-convolved from the data.  Consequently, this
acquisition mode does not allow sensible energy reconstruction
$\lesssim 100$~keV.  During zero-field data taking, which was used to
study the light collection properties of the detector, the PMTs were
operated at full gain, with sensitivity to single phes at 95\%
efficiency, and the data did not suffer from an effective low-energy
readout threshold.

\begin{figure}[htbp]
  \centerline{
      \includegraphics[width=1.0\hsize]{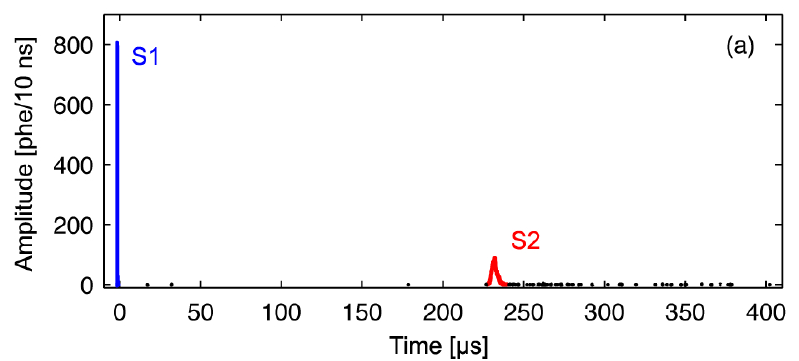}
  }
  \centerline{
      \includegraphics[width=0.5\hsize]{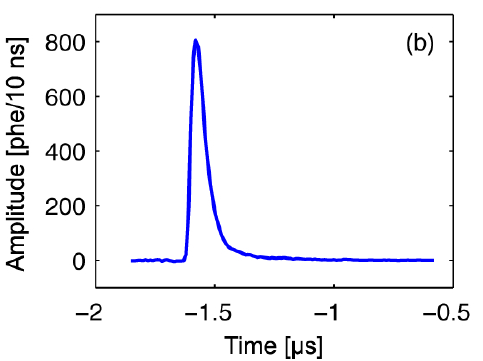}
      \includegraphics[width=0.5\hsize]{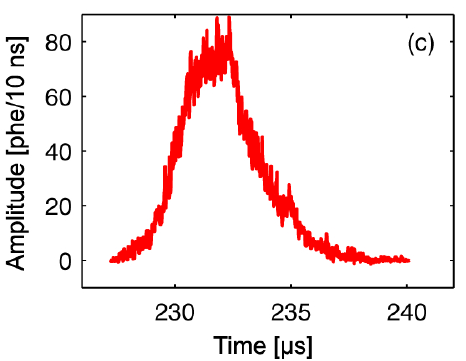}
  }
  \caption{Example dual-phase \textgamma~event at 28~cm depth from
    background data.  The entire event is shown in (a), while (b)
    shows a zoomed view of the S1~signal and (c) shows a zoomed view
    of the S2~signal.  The 49.5~cm length of the liquid xenon active
    region of the detector corresponds to 412~\ms~at our operating
    field of 62 V/cm.}
  \label{fig:sample_event}
\end{figure}

\begin{figure}[htbp]
  \centerline{
    \includegraphics[width=0.7\hsize]{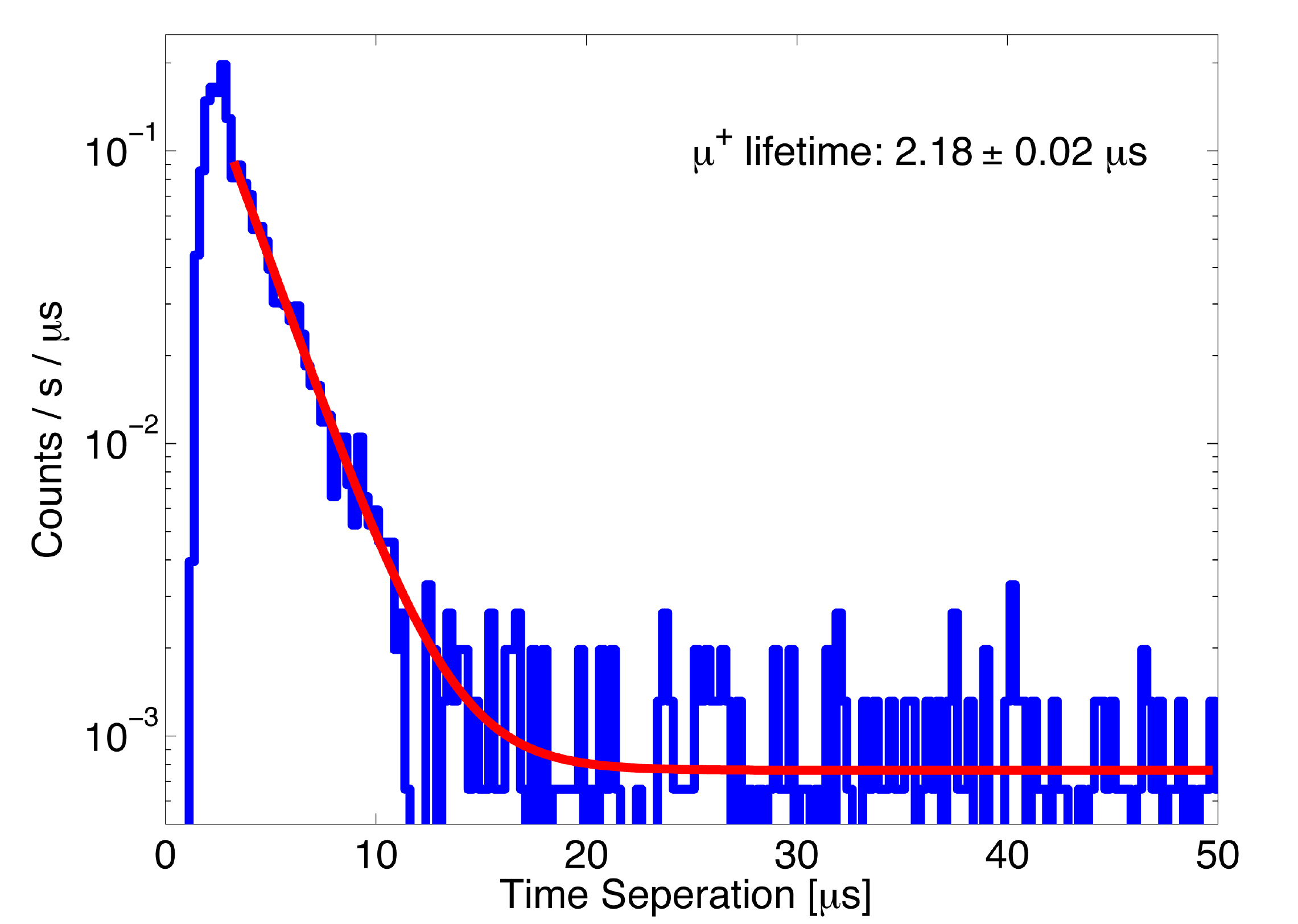}
  }
  \caption{Positively-charged muon lifetime measured in zero-field
    data.  We find agreement with the expected value of
    2.197~\ms~\cite{Ref:PDG2012}.}
  \label{fig:mu_life}
\end{figure}

Digitized pulse shapes are parameterized by a number of characteristic
quantities, such as pulse area, height, and length, that can be used
to identify S1 and S2~signals.  The LXe scintillation mechanism leads
to near-exponential VUV pulses with time constants of a few tens of
nanoseconds and very fast rise
times~\cite{Ref:Kubota1978,Ref:Kubota1979}.  Conversely, S2~signals
generated by electrons emitted from the liquid surface are typically
much larger and have durations of about 1~\ms, due to the drift time
of the electrons in the gas gap as well as the smearing of emission
times caused by electron diffusion in the liquid. These
characteristics allow a clear separation between S1 and S2~pulses.  A
typical event from the surface run is shown in
Fig.~\ref{fig:sample_event}.  We have checked the reconstruction of
the pulses and timing information with a number of basic measurements,
including the muon lifetime, shown in Fig.~\ref{fig:mu_life}.  This is
measured in the decay of $\mu^+\to\bar{\nu_{\mu}}{\nu_e}e^+$
using the measured time difference between an S1~signal characteristic
of a muon and a subsequent S1~signal.  This measurement was made in
zero-field data and did not take pile-up into account.  We find good
agreement between our measured value of $2.18\pm
0.02~\mbox{(stat.)}$~\ms~and the world average of
2.197~\ms~\cite{Ref:PDG2012}.

\subsection{Preparations for underground running}

During the surface commissioning, valuable experience in detector
assembly and deployment, operation, and data-taking was gained.  We
were also able to identify a handful of issues that we rectified prior
to the detector transport underground.  The inherent limitation in the
high-voltage delivery system was solved by a new feed-through that has
been developed and tested extensively at 100~kV.  A new cathode wire
plane with 0.5~cm wire spacing and a wire diameter of $206\pm 1$
$\mu$m has been assembled and installed.  Based on these improvements,
we aim to achieve 800~V/cm in the drift region during underground
operation, although 500~V/cm will provide adequate discrimination to
reach our sensitivity goal~\cite{Ref:Xenon100}.  We have upgraded the
circulation system to achieve even higher flow rates by ensuring that
all fittings are properly connected in the plumbing system and
implementing a series of checks that help to ensure the integrity of
the detector internal circulation lines prior to full detector
deployment.  Even without these additional improvements, the LUX
detector surface operation was very successful, demonstrating good
thermal control, a high gas flow rate through the purification system,
and excellent readout capability.

\section{Studies of the surface data} \label{sec:results}

During both zero-field and dual-phase data-taking, we collected a
number of useful datasets that allow us to study important properties
of the detector, such as the xenon purity, the light collection, and
the three-dimensional position reconstruction.  In the dual-phase data
we were able to use the \textgamma~background to measure a maximum
electron drift length spanning half of the active length of the
detector despite the non-optimal circulation path.  The internal xenon
circulation path was studied with a diagnostic $\Rn$ injection into
the detector, making use of the imaging properties of the TPC.  The
various backgrounds and radioactive sources used during surface
operation, such as the cosmic muon background,
cosmogenically-activated $^{129\rm{m}}$Xe and $^{131\rm{m}}$Xe
isotopes, the \textalpha~particles from the $\Rn$ decay chain, and an
external $\Cs$~source, provided a wealth of opportunities to study the
light collection properties of the detector and tune the simulation
developed for the LUX experiment.  Finally, we have employed the
$^{214}$Bi-$^{214}$Po near-coincident decays from the $\Rn$ decay chain to
study the resolution of the position reconstruction algorithms.

\subsection{Xenon circulation and purification} \label{sec:purity}

The removal of electronegative impurities is essential to allow the
electrons produced by ionization of the xenon atoms to drift away from
the interaction site and for the scattering event to be reconstructed
accurately along the 50~cm length of the active region of the LUX
detector.  Therefore, one of the goals of the surface commissioning of
LUX was to obtain good detector purity as an essential exercise of the
detector circulation systems in advance of the underground
data-taking.

\begin{figure}[tbhp]
  \centerline{
    \includegraphics[width=0.7\hsize]{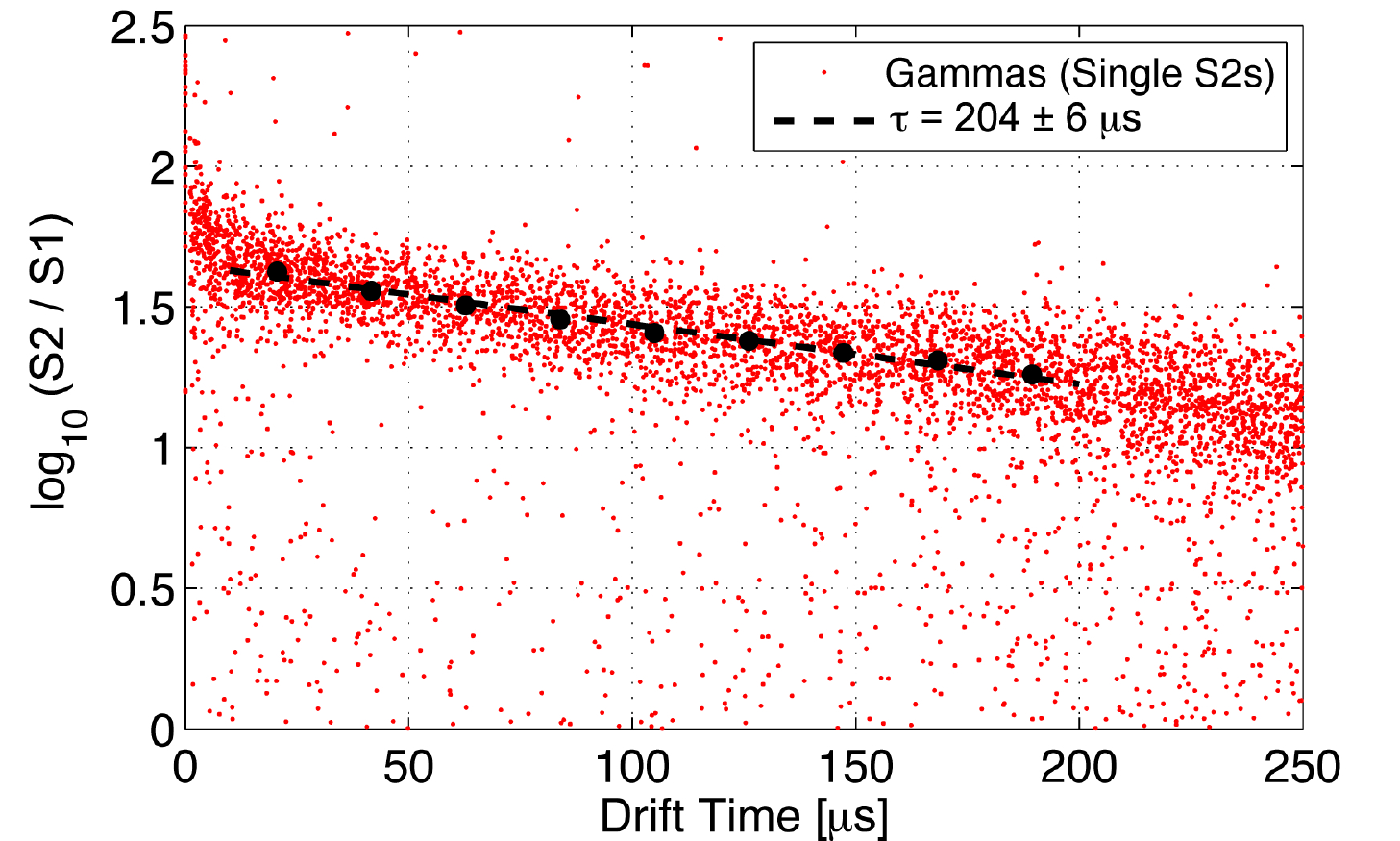}
  }
  \caption{Highest measured electron lifetime, obtained using
    dual-phase \textgamma-rays with a single pair of S1 and S2 signals
    in the event window. The circles indicate the means of the
    measured (S2/S1) ratio distribution in bins of drift time.  The S1
    signal in the event is corrected for observed depth dependence.}
  \label{fig:purity}
\end{figure}

We used dual-phase data to estimate the electron attenuation in the
active region of the detector.  The decrease in S2~signal size as a
function of drift time in the liquid, $\Delta t$, is described by an
exponential function $\mathrm{S2} = \overline{\mathrm{S2}} e^{-\Delta
  t/\tau}$, where $\tau$ is the mean lifetime that the electrons drift
in the xenon before being captured by electronegative impurities and
$\overline{\mathrm{S2}}$ is the size of the S2 signal in perfectly
pure xenon.  After exploring several modes of circulation, we obtained
an electron lifetime of $204 \pm 6$~\ms, which was measured using
background \textgamma~interactions with a single pair of S1 and
S2~signals in the event window.  In order to account for the fact that
the \textgamma's are not mono-energetic, we normalize each measured
S2~signal by its corresponding S1~signal, which was corrected for
observed depth dependence and x-y position dependence through light
response functions created for individual PMTs with the Mercury
algorithm~\cite{Ref:MercuryAlgo}, discussed further in
Sec.~\ref{sec:pos_reco}.  The resulting trend in (S2/S1), used to
obtain the electron lifetime, is shown in Fig.~\ref{fig:purity}. This
lifetime corresponds to a drift length of 25~cm.

\subsubsection{Imaging of the circulation path} \label{sec:rn_injection}

\begin{figure}[tbhp]
  \centerline{
    \includegraphics[width=0.7\hsize]{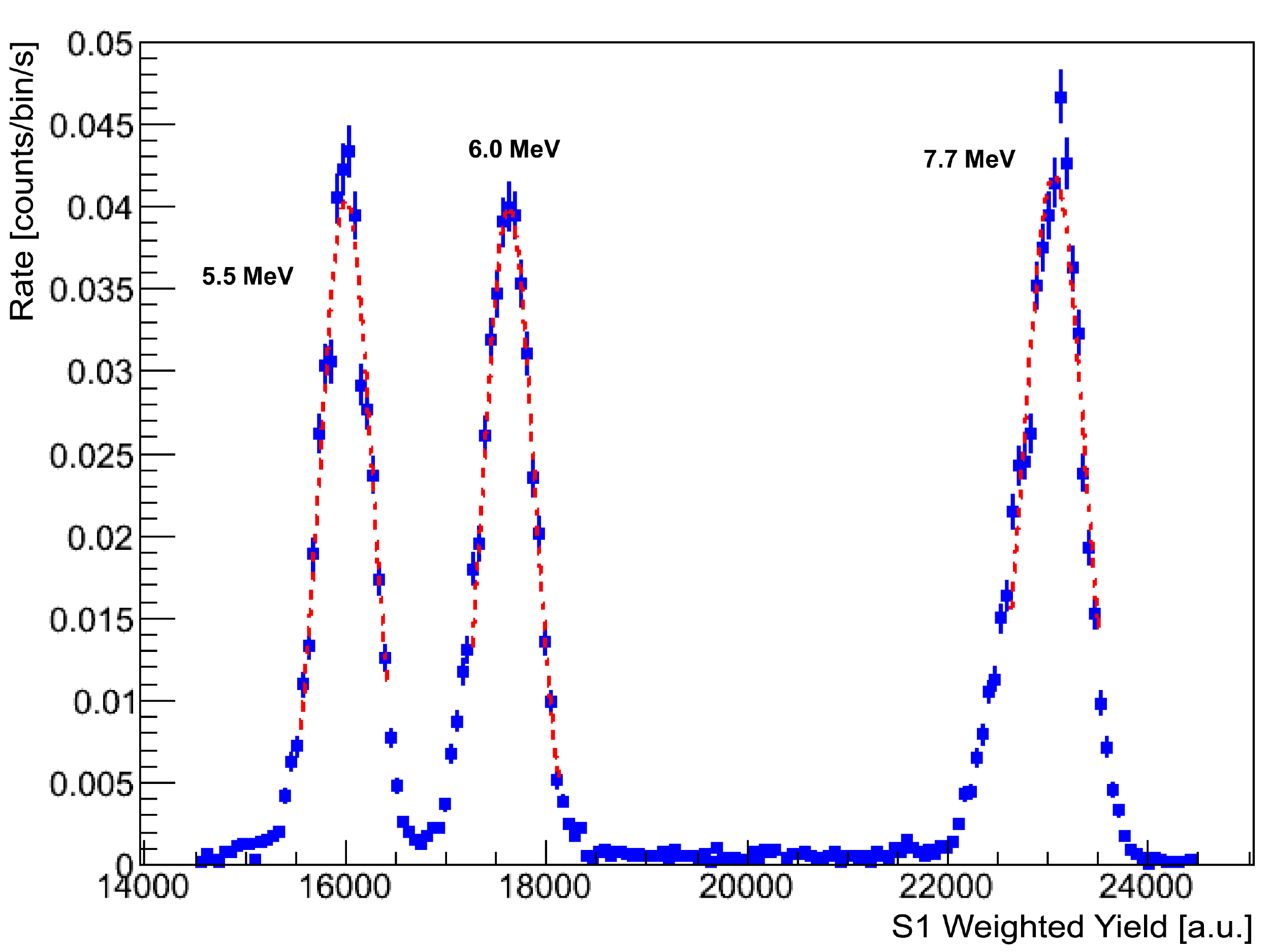}
  }
  \caption{S1 pulse areas [arbitrary units] of the three
    \textalpha~particles from the $\Rn$ decay chain for zero-field
    data.  The pulses are weighted to balance the light in the top and
    bottom PMTs, as described in Sec.~\ref{sec:light_collec}.}
  \label{fig:alphas}
\end{figure}

In order to verify the nature of the compromise in the circulation
path, 150~Bq of $\Rn$ were injected into the detector through a cold
trap to study the xenon flow, using the three \textalpha~particles
produced in the decays of $\Rn$ (5.5~MeV), $^{218}$Po (6.0~MeV), and
$^{214}$Po (7.7~MeV), shown in Fig.~\ref{fig:alphas}.  The $\Rn$ was
introduced into the system through a port in the circulation path that
is located before the getter so that the trace amount of impurities
introduced via the additional plumbing would be prevented from
reaching the detector. The $\Rn$ source had a steady-state emanation
rate of 1.42~Bq of $\Rn$ per minute of gas flow through the port.
Nitrogen was flown through the $\Rn$ source and into a cold trap that
captured the $\Rn$.  The amount of $\Rn$ captured into the trap was
controlled by the duration of nitrogen flow. Once the desired flow
time was reached, the nitrogen flow was stopped, the $\Rn$ source was
valved off and the cold trap with $\Rn$ was heated and introduced to
the circulation path.

\begin{figure}[tbhp]
  \centerline{
   \includegraphics[width=1.0\hsize]{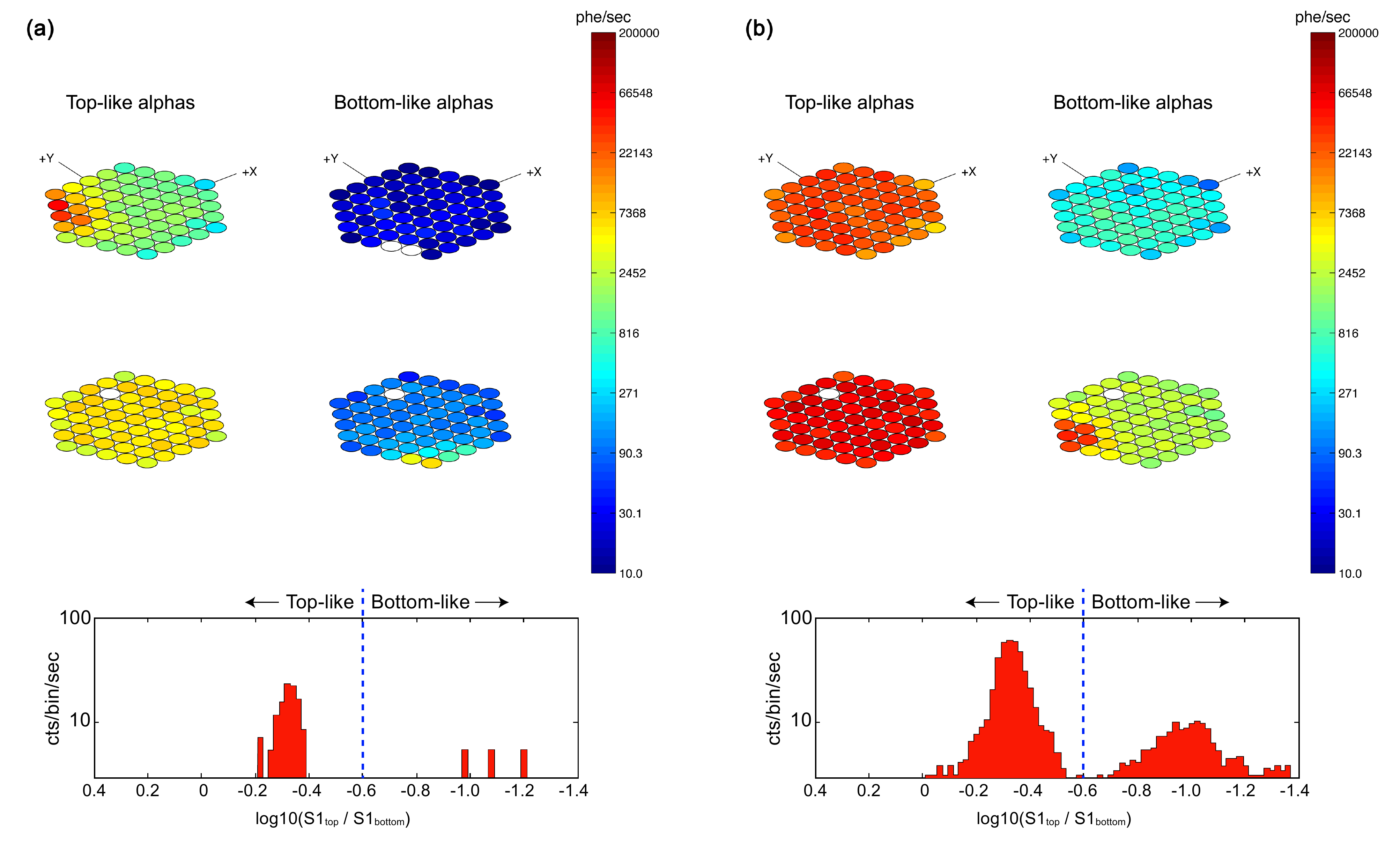}
  }
  \caption{The average zero-field S1~signal hit patterns and S1~light
    ratio for \textalpha~interactions in the LXe are shown at (a) 1.2
    minutes and (b) 17 minutes after the $\Rn$ injection.}
  \label{fig:rn_path}
\end{figure}

By studying the ratio of light observed in the top PMT array relative
to the bottom PMT array in zero-field data and the corresponding PMT
hit patterns, shown in Fig.~\ref{fig:rn_path}, we confirmed that the
$\Rn$~entered the detector in the second quadrant of the top PMT array
(left-hand plots of Fig.~\ref{fig:rn_path}).  After ten minutes, the
$\Rn$-doped xenon was seen entering the bottom PMT array at the same
location in the second quadrant, suggesting that the xenon was flowing
down this side of the detector (see the right-hand plots of
Fig.~\ref{fig:rn_path}).  This provided valuable diagnostic insight,
as the intended circulation path would have introduced xenon at the
bottom PMT array in the first quadrant instead.  Figure shows the
$\Rn$ flow from the top of the detector to the bottom, at two of the
time slices studied.  ``Top-like'' events pass the selection ${\rm
  S1}_{\rm top}/{\rm S1}_{\rm bottom} > -0.6$, the ratio of light in
the top PMT array relative to bottom PMT array, while events failing
this selection are considered ``bottom-like''.  Simulation indicates
that the S1~light relative asymmetry
$(\rm{S1}_{\rm top}-\rm{S1}_{\rm bottom})/(\rm{S1}_{\rm top}+\rm{S1}_{\rm bottom})$
provides a monotonic mapping to $z$-position, so this ratio is a good
proxy for depth.

It was expected and subsequently confirmed that this $\Rn$ injection
will not lead to significant backgrounds in a WIMP-search run. The
radon-related background could arise from $^{210}$Pb plate-out on the
PTFE~panels and other components, leading to low-energy interactions
into the liquid bulk, and from (\textalpha,n) neutron production on
fluorine in the PTFE; a particular concern is the mis-reconstruction
of the position of any surface interactions into the fiducial volume
(nominally located approximately 50~mm from the surface of the
panels).  The latter effect could lead to mis-reconstruction of
nuclear recoils from \textalpha~interactions and electron recoils from
\textbeta~interactions, both of which may suffer incomplete charge
extraction due to the proximity to the wall. Conservative calculations
for these processes indicate that 150~Bq of $\Rn$ will not compromise
the WIMP sensitivity. For example, if all $\Rn$ activity appeared as
$^{210}$Pb plated out onto the PTFE~panels, the total neutron
production rate from the (\textalpha,n) process would be 11~n/yr,
assuming $10^{-5}$~n/\textalpha~on a thick fluorine target.  This
corresponds to 7.5\% of neutron background expected from the PMTs,
which are predicted to contribute a mere 0.04~WIMP-like background
events in a nominal nuclear recoil search window of [5,25]~keV after
30,000~kg-days.

\subsection{Light collection} \label{sec:light_collec}

The light collection properties of the LUX detector are of utmost
importance for the dark matter search, as they affect directly the
size of the S1~signals and, consequently, our energy reach and
background discrimination.  In order to model the light collection
properties of the detector, we use zero-field calibration data, which
provides the best statistics in the S1~signal, to determine the number
of detected phe per keV deposited through electron recoils in the
xenon target.  Two particularly important aspects of the light studies
of the detector are the reflectivity of the PTFE panels in liquid
xenon and the photoabsorption length for the 178~nm VUV scintillation
photons.  Liquid xenon can be purified to be highly transparent to its
own scintillation light and, consequently, the attenuation length for
VUV photons depends on the xenon purity. This and other optical
properties of the chamber can be obtained by detailed comparisons
between real and simulated data.

The LUX detector is modeled with a GEANT4-based
simulation~\cite{Ref:GEANT,Ref:LUX_SIM}.  The simulation makes use of
the NEST~model~\cite{Ref:NEST}, which takes into account the energy-,
field-, and particle-dependent S1 and S2~signal yields, generating
signals with realistic means and resolutions.  While the detector
geometry has been implemented accurately from direct detector
measurements, a number of light collection properties of the detector
must be determined from data.  The unknown or uncertain parameters in
the simulation include the reflectivity of the PTFE surfaces, the
reflectivity of the wire planes, the photoabsorption length, the
Rayleigh scattering length, and the reflectivity of the aluminum
flashing deposited behind the quartz PMT windows.  These quantities
have been tuned separately for the gas and liquid regions of the
detector through extensive comparisons with $\Cs$ source data and then
validated using the other available data.  While the light yield has
traditionally been measured at 122~keV using the $\gamma$s from a
$^{57}$Co source, which have a 3~mm attenuation length in LXe, these
do not readily enter the fiducial volume in a detector the size of
LUX, so other calibration sources are necessary.

\begin{figure}[tbhp]
  \centerline{
    \includegraphics[width=0.7\hsize]{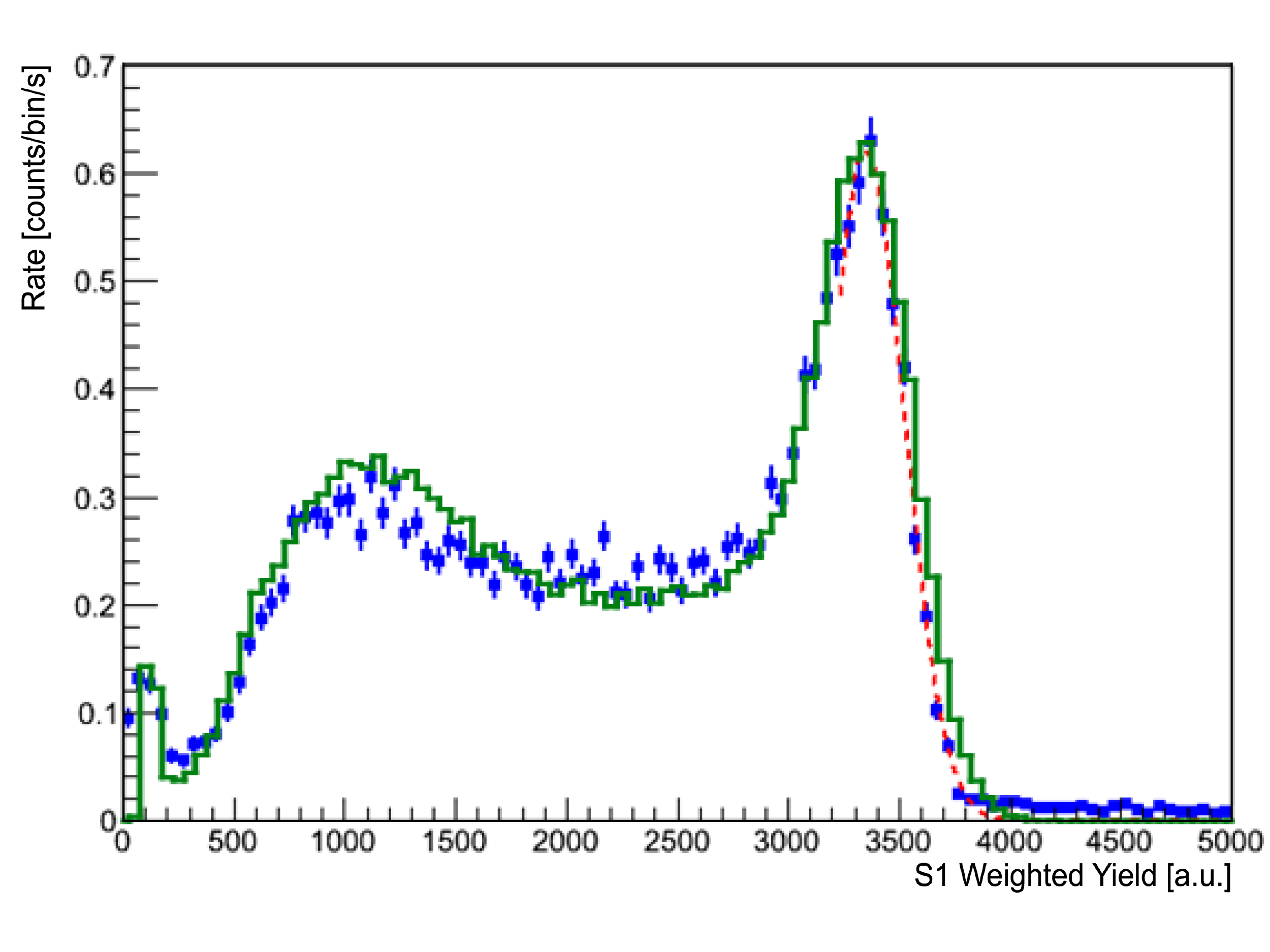}
  }
  \caption{$\Cs$ source S1~energy distribution [arbitrary units] for
    zero-field data.  The S1 signal is rotated to balance the light
    between distribution in the top and bottom PMTs and minimize
    energy resolution. The S1 signal from the tuned simulation
    (histogram) is compared with data (points with errors) collected
    using a $\Cs$ source located halfway down the length of the
    detector.}
  \label{fig:cs_energy}
\end{figure}

The full $\Cs$ energy deposition peak from the 662~keV \textgamma-rays
appears at $5624 \pm 8$~phe for a source located halfway down the
active region, corresponding to $8.4$~phe/keV, which is more than 2.5
times the $\approx$3~phe/keV value at zero-field reported for $\Cs$ in
Fig.~8 of Ref.~\cite{Ref:Xenon100}. To obtain the best resolution on
the 662~keV $\Cs$ full energy deposition peak, the S1~signal pulse
area measured in phe is weighted to correct for position-dependence of
the light collection by rotating the distribution between light
observed in the top and bottom PMTs to minimize the energy resolution.
Three sets of optical parameters are tuned for both gas and liquid
separately: the PTFE reflectivity, the photoabsorption length, and the
wire grid reflectivities. The light yield is determined from the data
and the simulation is tuned to match both the light yield observed in
data and the ratio of light in the top PMTs relative to the bottom
PMTs. The simulation is rotated in an identical way.
Figure~\ref{fig:cs_energy} shows that excellent agreement has been
achieved between the data and simulation for the $\Cs$ source, which
was contained in a 5~mm lead-backed collimator, lowered halfway down
the length of the active region in a tube next to the cryostat.  The
resolution in the simulation is shown without any additional scaling
or re-weighting, realistically replicating the resolution in data
using only the NEST~model.
Through the tuning of the optical properties of the simulation, we
conclude that the hemispherical reflectivity of the PTFE panels is $>
95\%$ (pure Lambertian) in LXe and the photoabsorption length is at
least 5~m in the liquid.  The best description of our data is obtained
with $100^{+0}_{-2}\%$ PTFE reflectivity and $11^{+2}_{-1}$~m
photoabsorption length in LXe.  The largest correlation between
optical parameters is 2\%.  An accurate determination of these values
is important for the thorough understanding of the LUX detector, as
well as for the design of future experiments, and we will repeat these
measurements in underground data-taking as well.

\begin{figure}[tbhp]
  \centerline{
    \includegraphics[width=0.7\hsize]{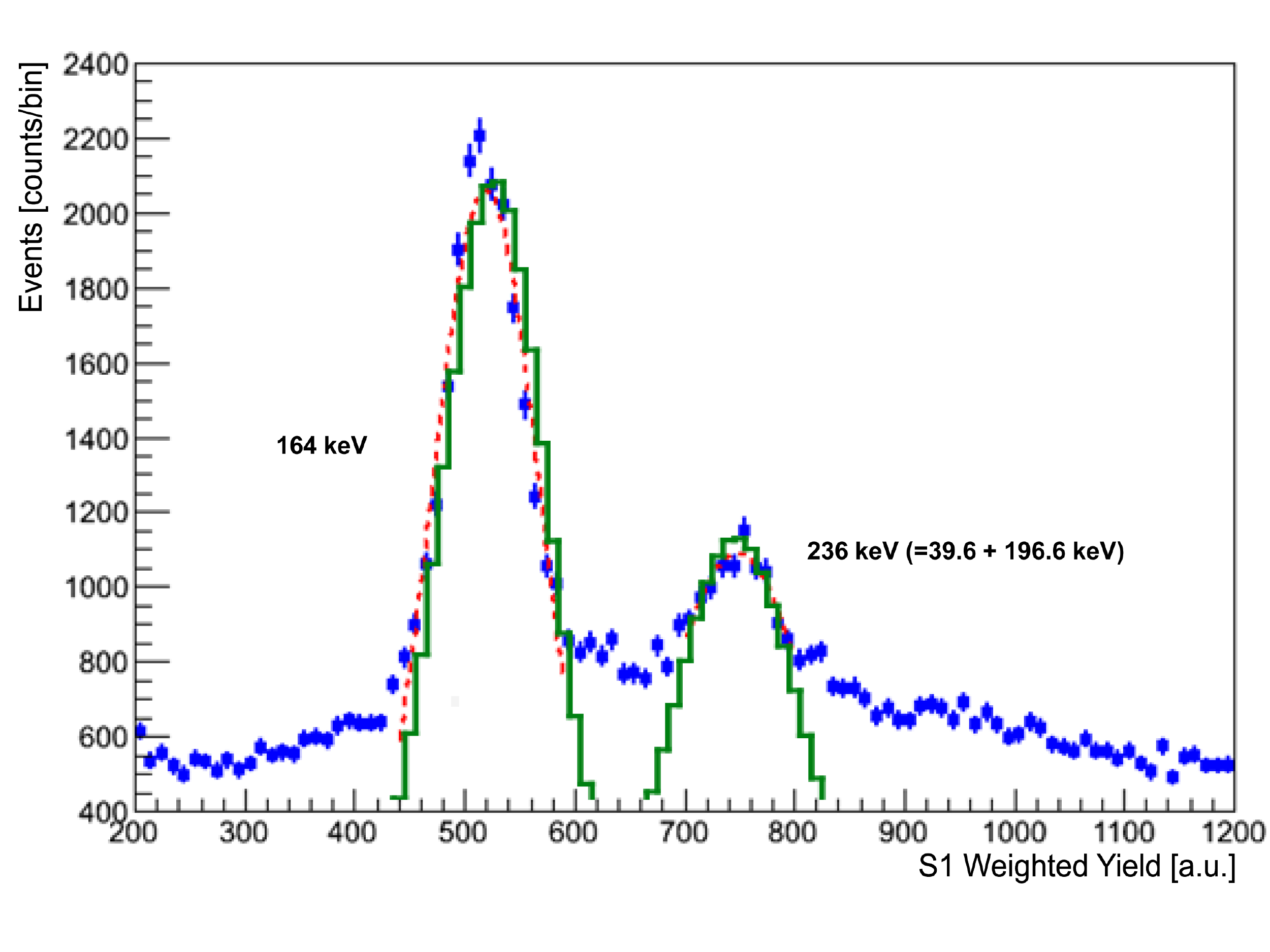}
  }
  \caption{Zero-field S1~signals from the \textgamma~decay of
    cosmogenically activated $^{129\rm{m}}$Xe (236~keV) and
    $^{131\rm{m}}$Xe (164~keV) (arbitrary units).  The resolution of
    the simulated S1 signal (histogram) agrees well with the data
    (points with error bars).}  \label{fig:activated_xe}
\end{figure}

In order to confirm the validity of the light model obtained from the
$\Cs$~data, we have compared the tuned simulation with the 236~keV and
164~keV \textgamma-rays from decays of $^{129\rm{m}}$Xe and
$^{131\rm{m}}$Xe, respectively.  We find good agreement in the
distribution of the data throughout the length of the detector,
determined through the asymmetry of light in the top PMTs relative to
the bottom PMTs, and in the energy resolution, shown in
Fig.~\ref{fig:activated_xe}, which lends confidence to our light
collection model.  The simulation again replicates the resolution
without additional scaling, even in the case of the non-monoenergetic
236~keV~\textgamma~emission from $^{129\rm{m}}$Xe, which consists of a
196~keV~\textgamma-ray~plus a second 40~keV~\textgamma-ray. Using the
simulation to inform the calculation of the volume-averaged light
yield gives $8.2\pm 0.2$~phe/keV for the xenon activation lines and
$7.8\pm 0.2$~phe/keV for the $\Cs$ source for a fiducial volume of
100--150~kg, where systematic uncertainties are indicated.

\begin{figure}[tbhp]
  \centerline{
    \includegraphics[width=0.7\hsize]{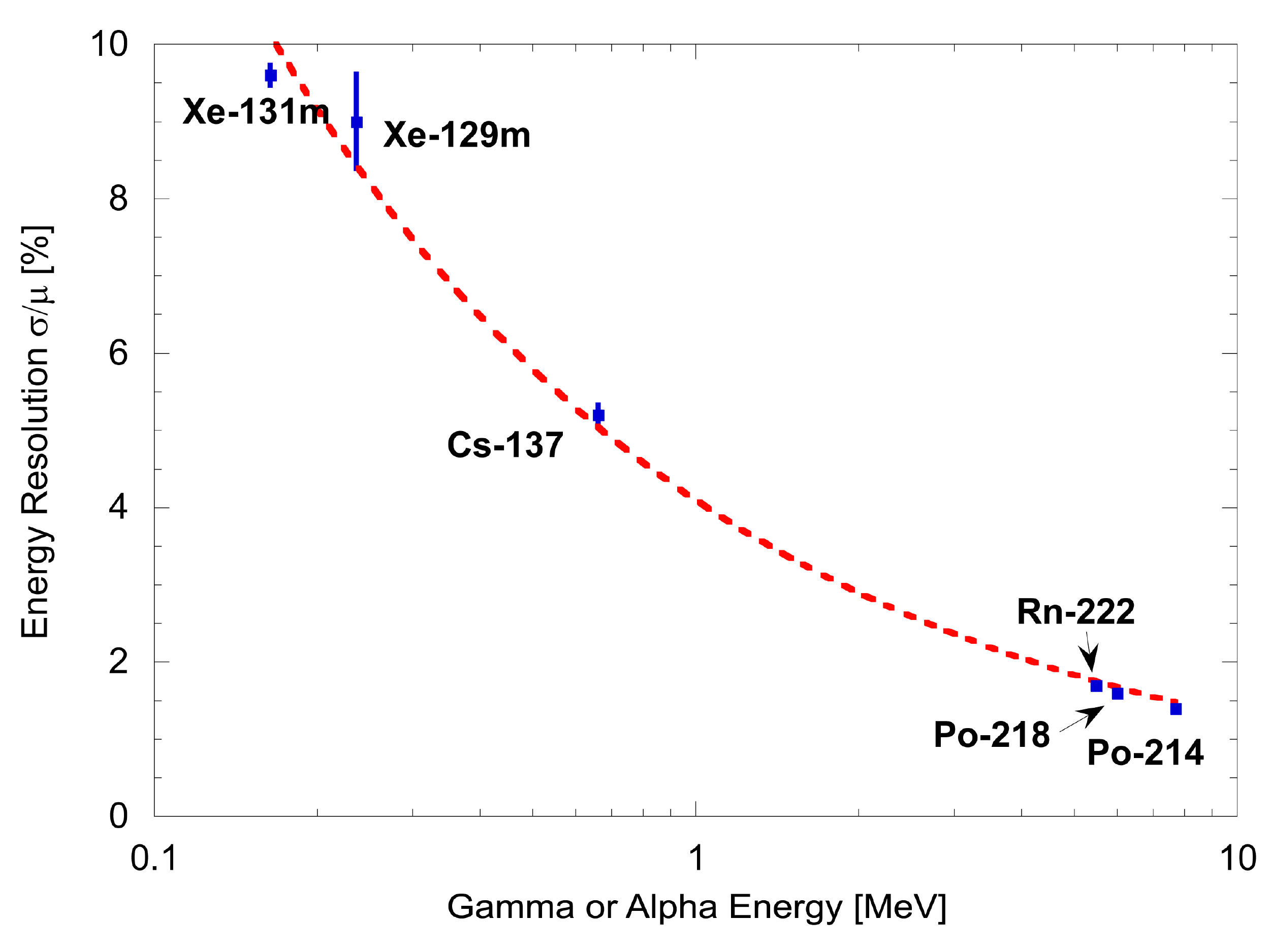}
  }
  \caption{The energy resolution $\sigma/E$ as a function of source
    energy for zero-field data.  The data points are shown with error
    bars.  We find a resolution of $\sim$\,2\% for the high-energy
    \textalpha~particles, while we observe a $\sim$\,10\% resolution
    for the 164~keV \textgamma-ray from the $^{131\rm{m}}$Xe decay.
    The trend in energy resolution is $\propto 1/\sqrt{E}$.}
  \label{fig:resolution}
\end{figure}

In addition to the excellent light collection and high PTFE
reflectivity we have observed, we also find good energy resolution at
our calibration energies.  We obtain $\sigma/E \sim$\,2\% for the
three \textalpha~particles, $\sim$\,5\% from the $\Cs$ full energy
deposition peak, and $\sim$\,10\% for the activated xenon
\textgamma-rays.  All resolutions are measured using zero-field data.
The energy resolutions as a function of energy are shown in
Fig.~\ref{fig:resolution}.  The resolutions have been obtained by
correcting the S1~signals for $z$ position-dependence.

\subsection{Position reconstruction} \label{sec:pos_reco}

\begin{figure}[tbhp]
  \centerline{
    \includegraphics[width=0.7\hsize]{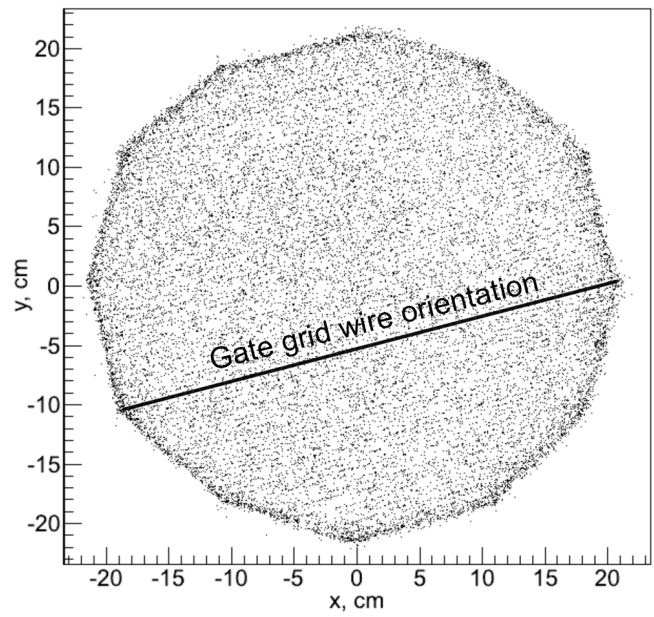}
  }

  \caption{The $x$-$y$ distribution of dual-phase background data
    reconstructed with the Mercury algorithm.  The gaps indicate the
    position of the gate wires, which are separated by 5~mm and
    inclined at an angle of about 20\degree.}
  \label{fig:gate_position}
\end{figure}

We have explored $x$-$y$ position reconstruction using the S2~signal
in dual-phase data with several algorithms, including basic centroid
methods.  We have also implemented the Mercury vertex reconstruction
algorithm~\cite{Ref:MercuryAlgo} developed for the ZEPLIN-III experiment, which provides
precise $x$-$y$ position information by measuring the light response
of each PMT {\it in situ}.  Figure~\ref{fig:gate_position} shows that
we are able to resolve the 5~mm wire spacing in the gate wire plane
when applying the Mercury algorithm to background data; the individual
wires appear as gaps because the electrons are focused around them in
our electric field configuration.  Owing to the lowered PMT gains, the
energy of these reconstructed background events is generally well
above that of the WIMP-search region, but the pulse sizes are nonetheless small.
\begin{figure}[tbhp]
  \centerline{
  \includegraphics[width=0.8\hsize]{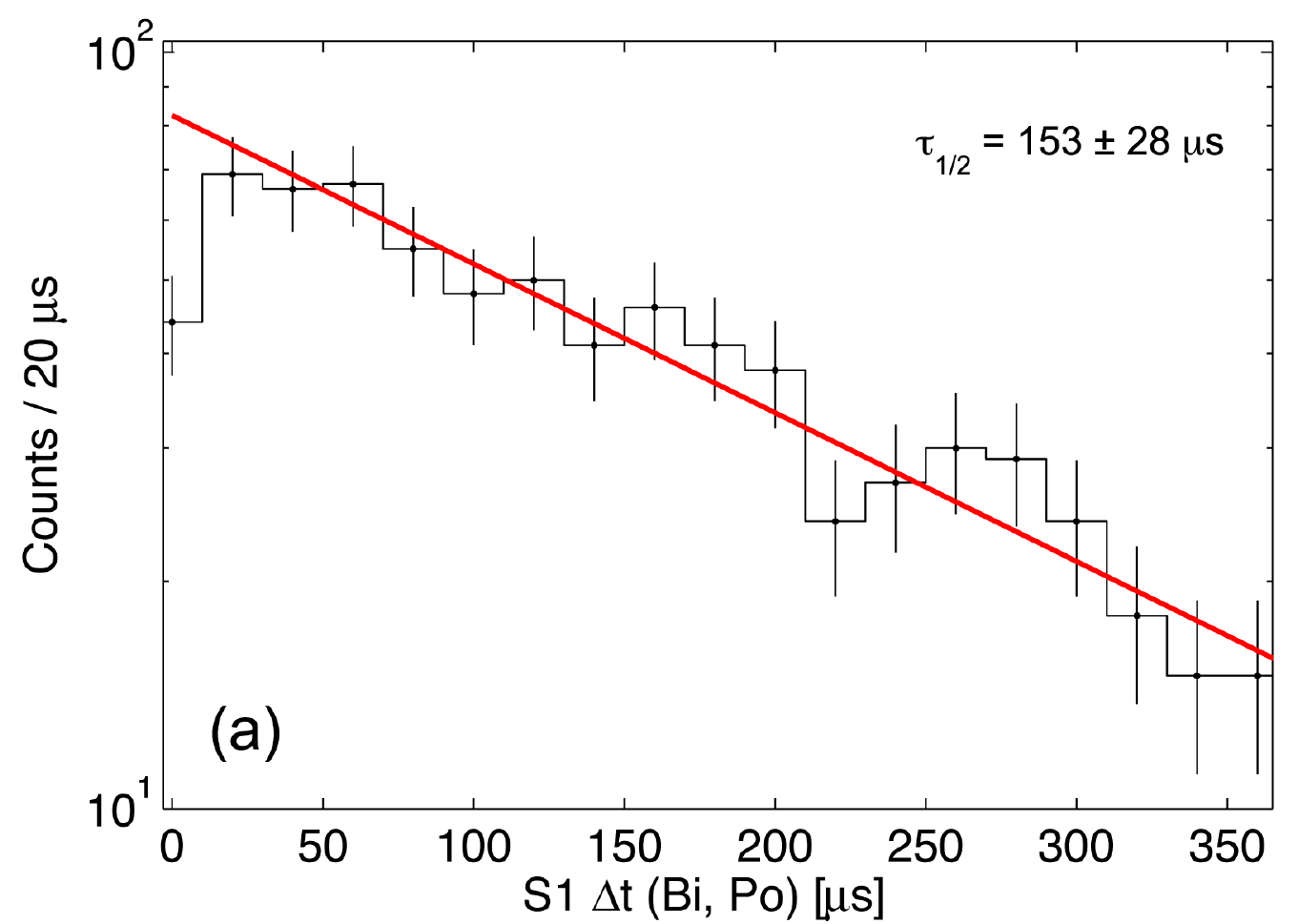}
  }
  \centerline{
      \includegraphics[width=0.4\hsize]{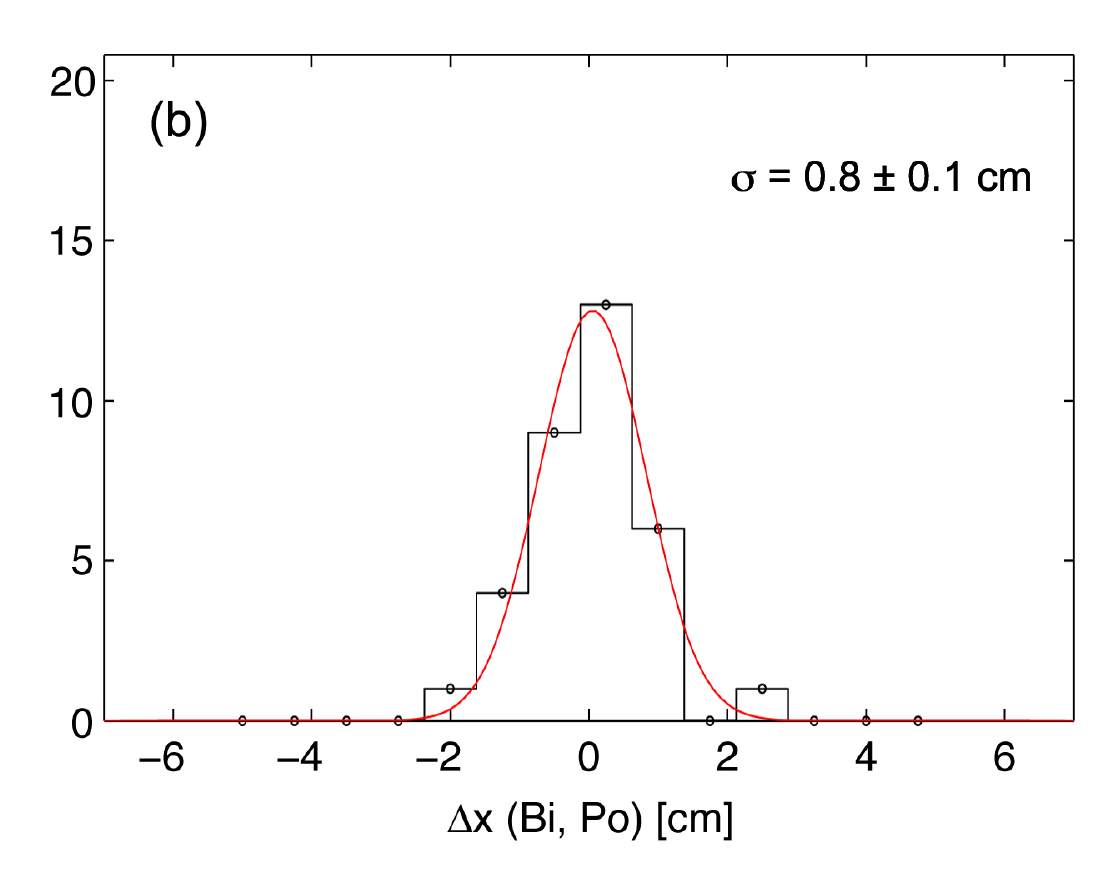}
      \includegraphics[width=0.4\hsize]{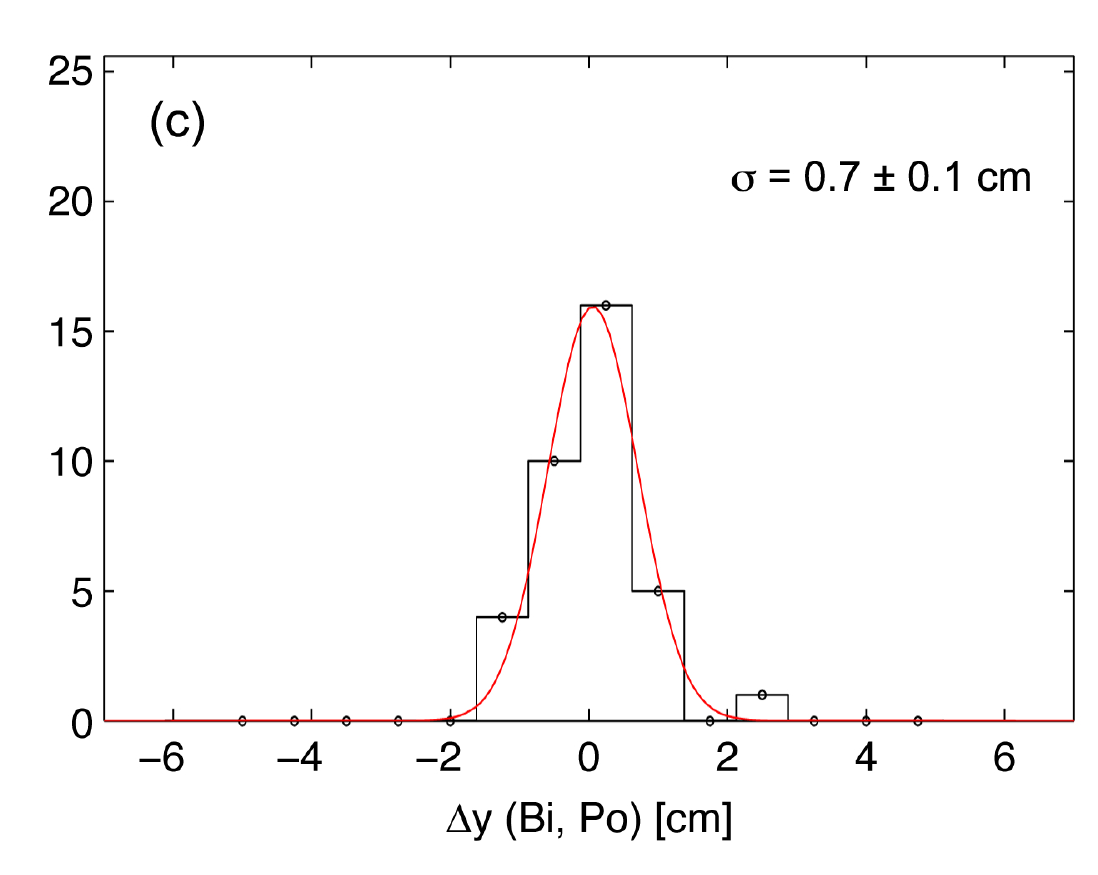}
  }
  \caption{The $^{214}$Bi-$^{214}$Po candidates' (a) half-life and
    statistical resolution along (b) the $x$-axis and (c) the $y$-axis
    of the detector for dual-phase data.  The reconstructed
    $^{214}$Bi-$^{214}$Po candidates show a half-life of $153\pm
    28$~\ms, which is consistent with the expected half-life of
    164.3~\ms~\cite{Ref:BiPo}.  The statistical component of the
    resolution in either lateral direction is measured to be
    $\sigma\sim$\,7~mm.}  \label{fig:bi_po}
\end{figure}

To study the statistical component of the position resolution using
the Mercury reconstruction, we have used the 7.7~MeV
\textalpha~particle emitted in the $^{214}\rm{Po}$ decay in delayed
coincidence with the \textbeta~decay of $^{214}\rm{Bi}$, which
precedes the $^{214}\rm{Po}$ decay with a half-life of
164.3~\ms~\cite{Ref:BiPo}.  Since our event window is 500~\ms~long, we
can search for these two decays in the same event.  In the
reconstruction we require a pair of S1 and S2 signals followed by a
second pair of S1 and S2 signals consistent with an
\textalpha~interaction.  Using these events, we measure a half-life of
$153\pm 28$~\ms, shown in Fig.~\ref{fig:bi_po}, which is in good
agreement with the known $^{214}\rm{Po}$ half-life of $164.3$~\ms~and
confirms that the selection has identified a viable sample of
$^{214}$Bi-$^{214}$Po coincident events.  We then use the relative $x$
and $y$ positions of the two decays to estimate the statistical
component of the resolution, finding $\sigma\sim$\,7~mm in either
direction for \textalpha~particles distributed in the bulk of the LXe.
This method treats the two sources as point-like, leading to an
overestimate in the quoted resolution as the \textbeta~track extends a
few millimeters in the liquid.  Because of the threshold imposed by
lowered PMT gains, these data were not taken in our nominal dark
matter operation mode.  Consequently, we estimate that the size of the
S2~signals is comparable to what we expect for our dark matter search,
which suggests a comparable resolution for our WIMP search.  However,
as this was an initial effort to convert the algorithm to LUX, we hope
it's performance will continue to improve with further study.  Since
the $\Rn$ decay daughters preferentially attach to the wire planes, we
have also observed a population of \textalpha~interactions from the
gate wire plane (not shown in the figure); these have a much larger
average S2~signal, as they are not suppressed by purity effects and
there is no appreciable lateral diffusion of the signal since they are
drifted only through the much higher extraction field in the LXe
before they reach the liquid surface.  These events have a resolution
of approximately 3~mm in both the $x$ and $y$ direction.

\section{Summary and Outlook} \label{sec:summary}

We have used the surface commissioning run of the LUX experiment to
exercise all xenon detector systems and have demonstrated operational
and readout capabilities.  The surface run has also helped us to
identify and address issues with the high-voltage delivery system, as
well as to implement thorough checkouts of all systems.  Benefiting
from the wealth of experience running and troubleshooting the detector
operations, we have developed confidence in our ability to deploy the
detector underground in a short period of time and to recommission it
for full operation.  We have taken advantage of the data collected to
study the behavior of the xenon purification dynamics, the light
collection of the detector, and to begin exploring event
reconstruction algorithms, as well as to tune and validate our
simulation using a variety of radioactive sources.

\begin{figure}[tbhp]
  \centerline{
    \includegraphics[width=0.7\hsize]{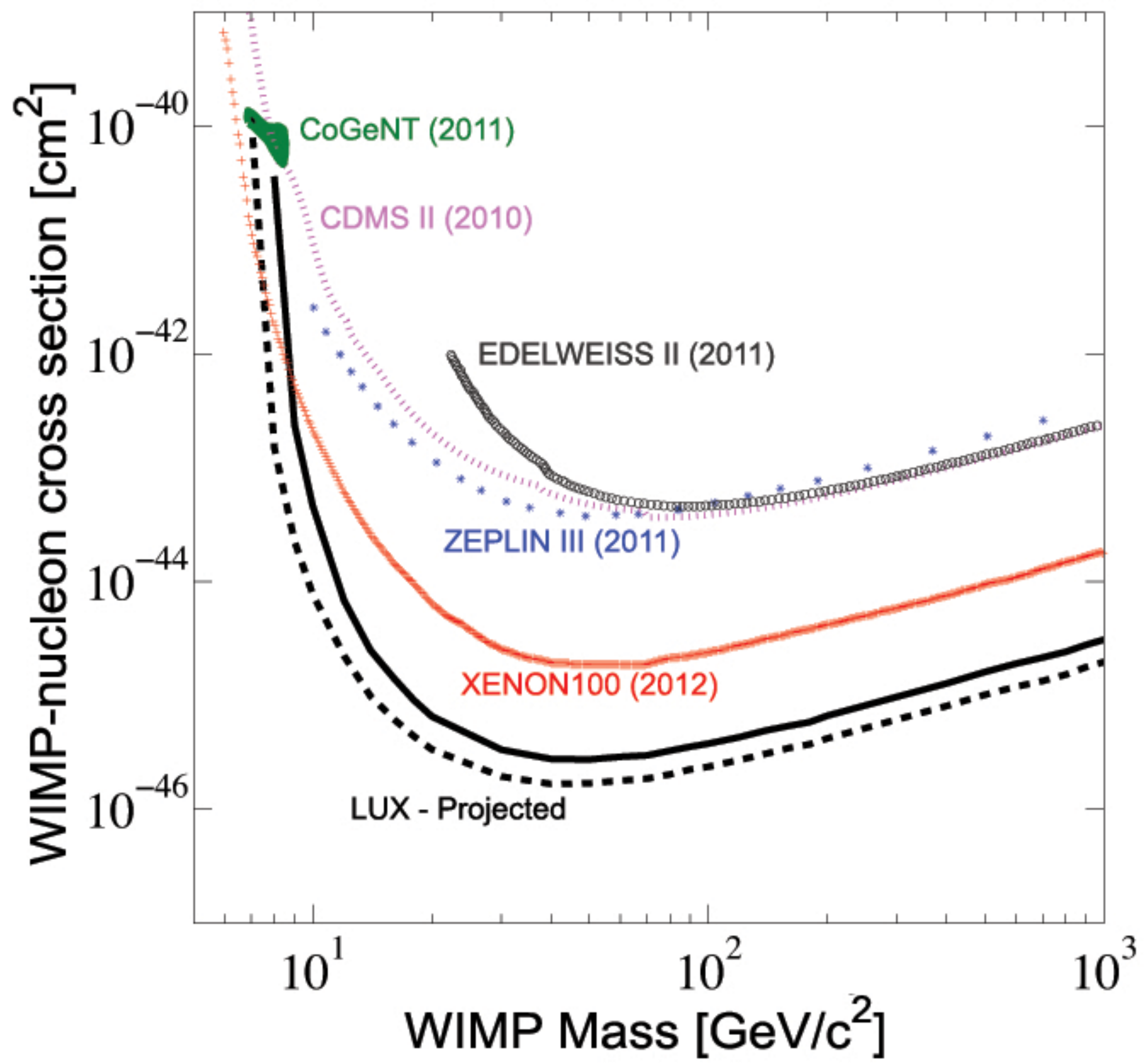}
  }
  \caption{The projected LUX sensitivity at 90\% CL is plotted in
  relation to other recent WIMP-nucleon scattering
  limits~\cite{Ref:Cogent,Ref:CDMSII,Ref:EdelweissII,Ref:ZeplinIII,Ref:Xenon100_new}.
  The solid black line shows a limit assuming very conservative light
  collection and 30,000~kg-days of data without background, while the
  dashed black line shows a realistic estimate of the limit given our
  current understanding of the light collection and 30,000~kg-days of
  data.}  \label{fig:sensitivity}
\end{figure}

Using the tuned simulation we have updated the WIMP sensitivity
estimates for 300 days $\times$ 100~kg of data assuming zero
background events and a minimum S1~signal threshold of 3~phe and an
operating field of 500~V/cm.  We have evaluated two scenarios in
estimating our sensitivity, shown in Fig.~\ref{fig:sensitivity}: a
conservative 15\% average photon collection efficiency, replicating
the state of the LUX detector at the end of the surface run, with a
WIMP search range of nuclear recoil energies between 4.4-25~keV and
50\% acceptance for nuclear recoils; and a more realistic light
collection efficiency of 20\%, which assumes full detector
purification and a WIMP search range of 3.5-25~keV as measured in
nuclear recoils and a 60\% acceptance for nuclear recoils.  The
nuclear recoil energy is determined from NEST, using the Hitachi model
for the Lindhard factor to obtain scaling of phe to nuclear recoil
energy.  The NEST model is in good agreement with the existing
$L_{eff}$ data~\cite{Ref:NEST}.  In both cases we expect sensitivity
at WIMP masses down to 10~GeV and in the realistic scenario we should
be able to make a definitive statement on the CoGeNT
result~\cite{Ref:Cogent} under standard dark matter halo assumptions,
assuming zero background events in the fiducial volume, and using
straight-forward analysis techniques that place a firm 3~phe minimum
threshold on the S1~signal and do not include lower energies that
statistically fluctuate upwards by setting the nuclear-recoil energy
scale to exactly zero at 3~keV, below which it is not possible to
compare the NEST model with
data~\cite{Ref:Manzur,Ref:Plante,Ref:Horn}.  Statistical fluctuations
in the recoil energy are included in the projection above the minimum
energy threshold.  We estimate a sensitivity to WIMP-nucleon cross
sections in the realistic scenario better than $2\times
10^{-46}$~cm$^2$ for a WIMP mass of 40~GeV/$c^2$, which exceeds the
original LUX sensitivity goal.

When we begin underground operation in late 2012, the LUX experiment
will become the largest two-phase xenon dark matter detector in the
world and will represent the first xenon TPC deployed in a water
shield.  During the underground operation of LUX, we expect the best
sensitivity to WIMP-nucleus scattering for WIMP masses above
8~\gevcc~in the realistic light collection scenario.

\section*{Acknowledgements}

This work was partially supported by the U.S. Department of Energy
(DOE) under award numbers DE-FG02-08ER41549, DE-FG02-91ER40688, DOE,
DE-FG02-95ER40917, DE-FG02-91ER40674, DE-FG02-11ER41738,
DE-FG02-11ER41751, DE-AC52-07NA27344, the U.S. National Science
Foundation under award numbers PHY-0750671, PHY-0801536, PHY-1004661,
PHY-1102470, PHY-1003660, the Research Corporation grant RA0350, the
Center for Ultra-low Background Experiments in the Dakotas (CUBED),
and the South Dakota School of Mines and Technology
(SDSMT). LIP-Coimbra acknowledges funding from Funda\c{c}\~{a}o para a
Ci\^{e}ncia e Tecnologia (FCT) through the project-grant
CERN/FP/123610/2011.  We gratefully acknowledge the logistical and
technical support and the access to laboratory infrastructure provided
to us by the Sanford Underground Research Facility (SURF) and its
personnel at Lead, South Dakota.


\bibliographystyle{elsarticle-num}
\bibliography{run2_paper}    


\end{document}